\documentclass{aa}

\usepackage{natbib}

\newcommand{\cd}{CD~$-24^\circ 731$}
\newcommand{\pg}{PG~1219$+$534}
\newcommand{\cpd}{CPD~$-64^\circ 481$}
\begin{document}

\title{Abundance studies of sdB stars using UV echelle HST/STIS 
spectroscopy\thanks{Based on observations made with the NASA/ESA Hubble
  Space Telescope, which is operated by the Association of
  Universities for Research in Astronomy, Inc., under NASA contract
  NAS 5-26555. These observations are associated with program
  \#8635.}}

\author{S.~J. O'Toole\thanks{\emph{Present address:}
Anglo-Australian Observatory, PO Box 296, Epping NSW 1710, Australia}
  \and U. Heber
}

\institute{Dr Remeis-Sternwarte, Astronomisches Institut der
  Universit\"at Erlangen-N\"urnberg, Sternwartstr. 7, Bamberg D-96049,
  Germany
}

\offprints{Simon O'Toole,
\email{otoole@aao.gov.au}}

\date{Received / Accepted}

\abstract
  {}
  {We test the hypothesis that the pulsations in sdB stars are correlated
    with the surface abundances of iron-group elements. Any
    correlation might explain why, when given two spectroscopically
    similar stars, one will pulsate while the other will not.}
  {We have obtained high-resolution ultraviolet
    spectra two pulsating and three non-pulsating sdB stars using the
    \emph{Space Telescope Imaging Spectrograph} onboard the
    \emph{Hubble Space Telescope}.  We determined abundances for 25
    elements including the iron group and even heavier elements such
    as tin and lead using LTE curve-of-growth and spectrum synthesis
    techniques.}
  {We find no clear correlation between pulsations and metal
    abundances, and we comment on the resulting implications,
    including whether it is possible to determine the difference
    between a pulsating and a non-pulsating sdB spectroscopically. In
    addition to the main goal of our observations, we have also
    investigated the effect of supersolar metallicity on fundamental
    parameter determination, possible trends with iron abundance, and
    the hypothesis that weak winds may be selectively removing
    elements from the stellar envelopes. These effects provide
    challenges to stellar atmosphere modelling and diffusion models
    for sdB stars.}
  {}

\authorrunning{S.~J. O'Toole \& U. Heber}
\titlerunning{UV spectroscopy of sdB stars}

\maketitle

\section{Introduction}
\label{sec:intro}

The subdwarf B (sdB) stars are core helium-burning objects with envelopes that
are too thin to sustain nuclear burning \citep[e.g.\ ][]{Heber86}. They can be
identified with models of Extreme Horizontal Branch (EHB) stars; in other
words, they have masses $\sim$0.5\,$M_\odot$ and will evolve directly to the
white dwarf cooling curve, bypassing the Asymptotic Giant Branch. While their
future evolution seems secure, the formation of these stars is uncertain. In
recent years radial velocity surveys have shown that a large fraction of sdBs
are in short-period binary systems \citep{MHMN01}, it appears certain that
binary interaction plays a significant role, however many objects are
apparently single. A binary population synthesis study by \citet{HanII} has
found that three channels can give rise to the observed characteristics of
sdBs: one or two phases of common envelope evolution; stable Roche lobe
overflow; and the merger of two helium-core white dwarfs. The latter scenario
could explain the population of single stars.

The possibility of pulsations in sdB stars was theoretically predicted by
\citet{CFB96} at around the same time they were observed by
\citet{ECpaperI}. The more than 30 known pulsators (officially known as
V361\,Hya stars) have $T_{\mathrm{eff}}=29\,000-35\,000$\,K and
$\log g=5.2-6.0$, periods of 1-10 minutes and amplitudes less than 60\,mmag
\cite[see ][for a review]{Kilkenny2002}. The driving mechanism of the
oscillations is believed to be related to the ionisation of iron and other
heavy elements at the base of the photosphere \citep{CFB97a}. As is the case
for other types of pulsators \citep[e.g.\ the PG~1159 stars,][]{QFB04} there is
an overlap in the ($T_{\mathrm{eff}}, \log g$) plane between pulsators and
non-pulsators \citep{ECpaperXII}. Diffusion calculations by \citet{CFB97a}
suggest that the surface iron abundance of pulsators should be higher than
that of non-pulsators, however studies by \citet{EHN06}, \citet{HE04} and
\citet[][ hereafter HRW]{HRW00} find that iron has approximately solar
abundance in most sdBs.

For this reason we set out to determine if any correlation exists between
surface abundances of iron-group elements for pulsators and
non-pulsators. Since elements such as nickel, manganese and chromium are not
normally accessible through ground-based optical spectra, it was necessary to
acquire high-resolution UV echelle spectra with the \emph{Space Telescope
  Imaging Spectrograph} onboard the \emph{Hubble Space Telescope}
(\emph{HST/STIS}). If it is not possible to separate
groups based on abundances, perhaps differing mass-loss rates contribute
significantly, as suggested by \citet{FC97}. Regardless of this, our abundance
measurements will be extremely useful for testing diffusion theory. Previous
studies of sdBs using UV spectra from the \emph{International Ultraviolet
  Explorer} (\emph{IUE}) suffered from mediocre S/N as well as
from poor, or a lack of, atomic data; this was especially the case for the
iron group \citep[e.g.\ ][]{BHS82,BSK82}.

\begin{table*}
\caption{Observations of six sdB stars using \emph{HST/STIS}. Note that
  \cpd\ was observed only in the FUV and using the E140H grating.}
\label{tab:obs}
\begin{center}
\begin{tabular*}{1.00\textwidth}{l|cccccccccc}
Target & $\alpha$ & $\delta$ & $V$ & $T_{\mathrm{eff}}$ & $\log g$ & log(He/H)
& Ref. & Pulsator? & $T_{exp}^{NUV}$ & $T_{exp}^{FUV}$ \\
       & (2000)   & (2000)   & (mag) & (K) & & & & & (s) & (s) \\
\hline
\pg & 12 21 29.1 & +53 04 37 & 13.2 & 33\,500 & 5.87 & $-$1.6
& 4 & yes & 3100/3100 & 2160/3100 \\
Feige~48     & 11 47 14.5 & +61 15 32 & 13.5 & 29\,500 & 5.54 & $-$2.9
& 2 & yes & 3500/3511 & 3600/3600 \\
\cpd & 05 47 59.3 & -64 23 03 & 11.3 & 27\,500 & 5.60 & $-$2.5 & 3 & no & --- &
1440 \\
Feige~66     & 12 37 23.5 & +25 03 60 & 10.6 & 34\,500 & 5.83 & $-$1.6 & 4
& no & 824 & 875 \\
\cd       & 01 43 48.6 & $-$24 05 10 & 11.0 & 35\,400  & 5.90 & $-$2.9 & 4
& no & 800 & 878 \\
\hline
\end{tabular*}
\end{center}
References: (1) \citet{HRW99}; (2) \citet{HRW00}; (3) \citet{OJF05}; (4) This
work.
\end{table*}

The  two pulsators we have chosen to observe are 
Feige~48 \citep{ECpaperXI},  and
\pg\ \citep{ECpaperXII}. 
Quantitative analysis of the first two objects using optical
spectra was carried out by HRW and the latter by \citet{HRW99}. As comparison
objects we have chosen Feige~66 and \cd\ (alias SB~707); a high-resolution \emph{IUE}
spectrum of the former was analysed by \citet{BHS82}. A very high-resolution
spectrum of the sdB \cpd\ was found in the \emph{HST} archive;
since its temperature is close to that of Feige~48, and it is not pulsating
(Koen, private communication), we decided to use it as a comparison
star to Feige~48. Note that these two stars, along with \cd, are in
close binaries. Feige~48 and \cpd\ have periods of 0.376\,d and
0.2772\,d respectively, while \cd\ has a period of 5.85\,d
\citep{OHB04,EHA05}. The companions of Feige~48 and \cd\ are both
most likely white dwarfs, while the nature of the companion of
\cpd\ is uncertain.

In this paper we present a detailed abundance analysis of each of these
objects
based on UV echelle
spectra obtained using \emph{HST/STIS}. We discuss the possible
correlation between iron-group abundances and pulsation, the
abundances of heavy elements in the context of radiative acceleration
and the trends discussed by \citet{SJOT04}, and
a solution to the temperature discrepancy seen between Balmer line fitting and
helium ionisation equilibrium.

\section{Observations}
\label{sec:obs}

Observations were made using the \emph{Space Telescope Imaging Spectrograph}
(\emph{STIS}) onboard the \emph{Hubble Space Telescope} (\emph{HST}). We
observed five stars, three pulsators and two non-pulsators for comparison. One
of the non-pulsators, Feige 66 is a spectroscopic twin of one of the pulsator
\pg. Details of the observations are shown in Table \ref{tab:obs}.
The stars were observed in the near- and far-UV (NUV and FUV) using the medium
resolution echelle E140M and E230M grisms. The NUV spectra each have a central
wavelength of 1978\,\AA, covering the wavelength range 1700-2370\,\AA, while
the FUV spectra are centered on 1425\,\AA, covering the 1160-1730\,\AA\
range. For all spectra a slit width of 0.2''$\times$0.06'' was
used. Each pulsator was observed in time-tag mode, where the
arrival times of each photon is recorded; the two non-pulsators were observed
in histogram mode. The spectra of Feige 48 have yielded one piece of
serendipity -- the discovery of velocity variations indicating a binary
companion to the star \citep{OHB04}.
Additionally, we found very high-resolution spectra
of Feige~66 and \cpd, taken using the E140H grism with a
0.2''$\times$0.2'' slit covering the range 1163-1363\,\AA. All of the
spectra are sharp-lined as expected from the low projected rotation
velocities seen in sdBs \citep[][and HRW]{OHB04}.

\section{Synthetic spectra}
\label{sec:synth}
In hot subdwarfs, most of the spectral lines due to iron-group
elements (V, Cr, Mn, Fe, Co, Ni) are found blueward of $\sim$2300\,\AA,
and in many cases continuum definition is difficult at the resolution
of our spectra. The methods we used to get around these problems are
discussed in more detail in Section \ref{sub:contdef}.

As input for our spectrum synthesis, we used a metal line-blanketed LTE model
atmosphere with solar metalicity and Kurucz' ATLAS6 Opacity Distribution
Functions. The spectra were synthesised using Michael Lemke's version of the
LINFOR program (originally developed by Holweger, Steffen, and
Steenbock at Kiel University). Oscillator strengths were taken from
the Kurucz line list, as were damping constants for all metal
lines. Only lines that have been observed experimentally were used, since we
required the most accurate wavelengths possible. In the case of species
heavier than Zn, values were taken from the resonance line lists
of \citet{Morton2000,Morton2003}. For the partition functions of
Ga\,\textsc{iii}, Ge\,\textsc{iv}, Sn\,\textsc{iv} and Pb\,\textsc{iv}
we used these ions' ground state statistical weight, since no
published data is available. This is a good approximation at
temperatures of 30\,000-35\,000\,K. The Cu\,\textsc{iii} atomic data were
taken from the tables of \citet{HH95}. Oscillator strengths for
Ti\,\textsc{iii} lines were taken from \citet{RU97}. Lines
with $\lambda>2000$\,\AA\ were converted from air to vacuum wavelengths
using the formula of \citet{Edlen1966}.

\subsection{Atmospheric parameters}
\label{sec:atmospar}

The determination of atmospheric parameters for the two pulsators to
be discussed here was presented by HRW (the values are shown in Table
\ref{tab:obs}).
We selected Feige~66 and \cd\ as potential comparison stars for the
pulsator \pg\ because they are not known to pulsate and
published temperatures and gravities are similar to those of the latter.
While \pg\ has $T_{\mathrm{eff}}=33\,500$\,K, $\log g=5.85$
from Balmer line analysis by HRW, Feige~66 has
$T_{\mathrm{eff}}=33\,400$\,K, $\log g=6.2$ \citep{SBK94}
and \cd\ has $T_{\mathrm{eff}}=34\,000$\,K, $\log g=6.0$ \citep{HHJ84}.

For this work we have reanalysed the Keck-HIRES spectrum of \pg\ from
HRW. A high resolution (0.1\AA) optical spectrum of \cd\ was kindly
provided by M.\ Altmann and H.\ Edelmann. It was taken with the FEROS
spectrograph at the ESO 2.2m telescope. A low resolution (5\AA)
spectrum of Feige~66 taken with the CAFOS spectrograph was provided by
M.\ Altmann. The spectral analysis of both spectra
is described below. The parameters for \cpd\ have already been
determined by \citet{OJF05}.

\subsection{The hot sdB temperature discrepancy}
\label{sub:tempfix}

As has been noted by HRW, there is a discrepancy between temperatures
derived from Balmer line fitting and helium ionisation equilibrium. For
\pg, the difference in $T_{\mathrm{eff}}$ is 2000\,K.

After the initial discovery of strongly supersolar abundances of heavy
metals  in three of the programme stars (Feige 66, \cd\ \&
\pg), we  investigated how this affects the determination of
the atmospheric  parameters. HRW used metal-line blanketed models with
solar metal content. Using the same model
atmospheres described above but with metals scaled by a factor of 10
([M/H]=1.0), we have recalculated both our abundances and stellar
parameters ($T_{\mathrm{eff}}, \log g$). There are currently no
opacity distribution functions with higher metalicities. Also,
despite the fact that iron itself is show roughly solar abundances for
all our targets, in many cases the nickel abundances is 1-2 dex above
solar. It is possible that in these cases nickel may become the
dominant source of line opacity.

\begin{figure}
\vspace{9cm}
\begin{center}
    \includegraphics{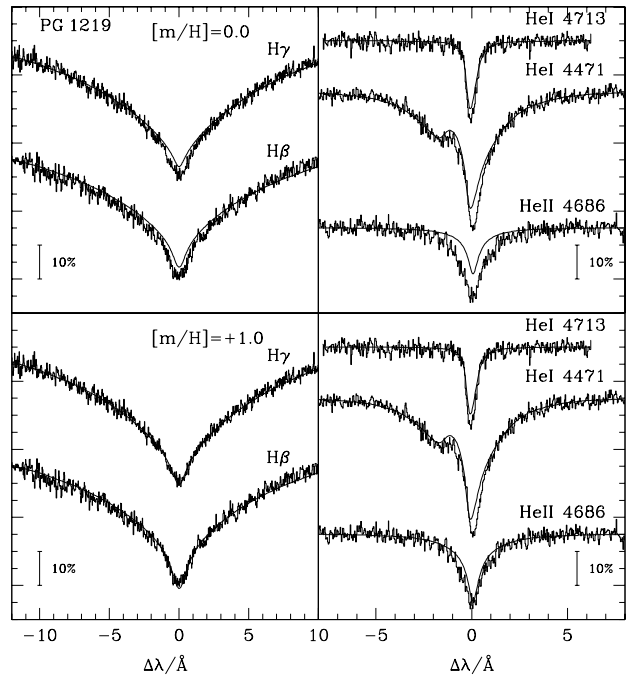}
\end{center}
\caption{Line profile fit for \pg\ using solar metalicity
  models (\emph{top panel}) and metal-rich models (10 times solar,
  \emph{bottom panels}). The Balmer lines and He\,\textsc{ii} at
  4686\,\AA\ match simultaneously when using the metal rich models
  (see text).}
\label{fig:pgp1fit}
\end{figure}

In Figure \ref{fig:pgp1fit} we show a fit with these models to the
optical spectrum of \pg. The Balmer lines and the
He\,\textsc{ii} can now be matched simultaneously. The parameters we
derive with our metal-enhanced models using all lines are closest
(within errors) to those found by
fitting the Balmer lines only with a solar-metalicity model. The
He\,\textsc{i} lines agree reasonably well, although the line cores of
the strongest lines match poorly. We expect ultimately the solution
will be found using opacity sampled models that can allow for enhanced
abundances relative to iron.

\begin{figure}
\vspace{9cm}
\begin{center}
    \includegraphics{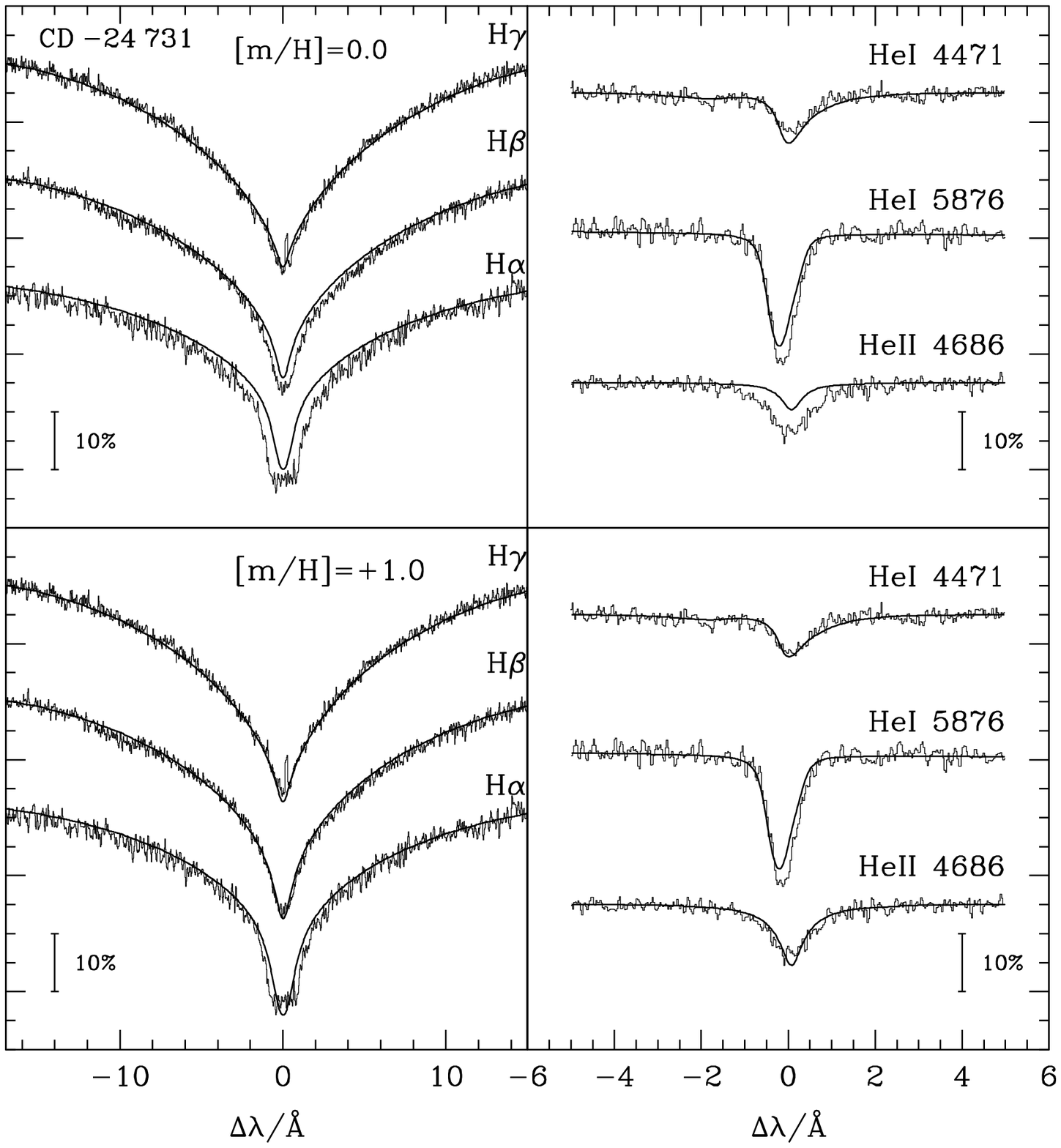}
\end{center}
\caption{Same as Figure \ref{fig:pgp1fit} but for \cd.}
\label{fig:p1fit}
\end{figure}

For \cd\ we found the same mismatch of the helium ionisation equilibrium
as for \pg\ when solar metalicity models were used (see Figure
\ref{fig:p1fit}).  Again this
problem is remedied by using metal-enriched models. In this case
a significantly higher effective temperature was derived from
the metal-enriched models than from the solar metalicity ones,
i.e. $T_{\mathrm{eff}}=35\,400$\,K compared with
$T_{\mathrm{eff}}=33\,800$\,K. \cd\ is a rather helium poor star
compared to Feige~66 and \pg, which might explain the different
metallicity dependence of the results. In addition, \cd\ resides in a
close binary system ($P$=5.85\,d), whereas Feige~66
is not known to be variable (E. Green, private communication). \pg\ is
also not known to be a radial velocity variable.

The low resolution spectrum of Feige~66 was analysed in the same way with
the metal-enriched models and we derived $T_{\mathrm{eff}}=34500$\,K,
$\log g=5.83$, log(He/H)=$-$1.6. In this case the parameters
agree with those found using non-LTE models; however, the He\,\textsc{ii}
4686\AA\ line does not agree with observations in the enhanced
metallicity case, but does in the non-LTE case. A high-resolution
spectrum is urgently required to resolve this issue.

The results of the new spectral analyses of \pg, \cd\ and Feige~66
revealed that their surface gravities are identical. The helium abundances
of \pg\ and Feige~66 are also identical, whereas helium abundance of
\cd\ is much lower than for the other two.
Their effective temperatures, however, are not identical as suggested by
the earlier spectral analyses. Feige~66 and \cd\ are hotter than
\pg\ by 1000K and 1900K, respectively. 

The temperatures we derive here match those determined using metal-free,
non-LTE models by HRW. Recently, \citet{CFB05} used \textsc{Tlusty} and
\textsc{Synspec} NLTE models and confirm the NLTE results of HRW for
\pg\ to within 500\,K and 0.04 dex.
Since the heavy metal abundances of Feige~48 and
\cpd\ are much closer to solar than those of the former, a
reanalysis of these stars is not necessary. The abundances shown in
Table \ref{tab:4abu} are derived from models with these parameters.

\section{Line identification}
\label{sec:lineID}

As has already been noted, previous studies of high resolution UV
spectra of hot subdwarfs have suffered from poor, or a lack of, atomic
data. While the current situation is by no means perfect, we have
endeavoured to identify as many spectral lines as possible, and then
from these we have chosen suitable lines to measure
equivalent widths or carry out spectrum fitting. Our method is outlined
below, beginning with a brief discussion of the problems encountered
with continuum definition.

\subsection{Defining the continuum}
\label{sub:contdef}

\begin{figure}
\vspace{6cm}
\begin{center}
    \includegraphics{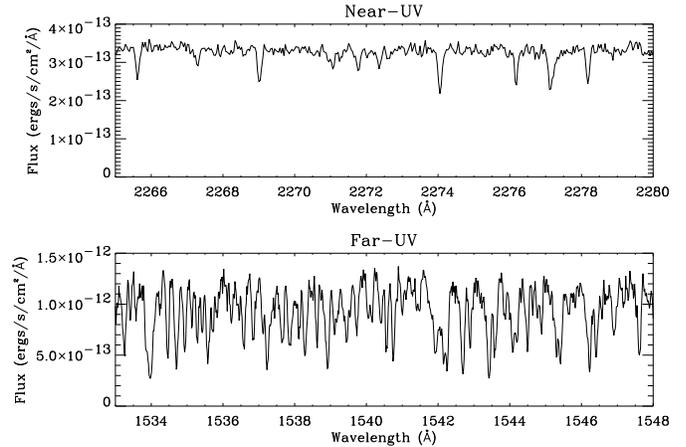}
  \end{center}
  \caption{Comparison of a 15\,\AA\ wide section of the near- and far-UV
  spectrum of \pg. In the NUV it is possible to measure the continuum
  level, while in the FUV it must be estimated. Note also the higher flux in
  the FUV.}
  \label{fig:compspec}
\end{figure}

In each of our spectra, the continuum is well-defined in the NUV,
especially redward of 2000\,\AA, allowing equivalent widths to be
measured. An example of the difference between crowded and
well-defined is shown in Figure \ref{fig:compspec}. There are also some regions
below 2000\,\AA\ where equivalent width measurements are possible. In
order to define the continuum in crowded regions, we measured
equivalent widths of all unblended lines and then generated synthetic
spectra using the abundances derived from this analysis; the continuum
of the crowded regions was set by matching the synthetic spectra to
the observations using a ``chi-by-eye'' fit. This was not always
straightforward however, since many lines still suffer from imprecise
atomic data, and some resonance line profiles are not well reproduced
in our assumed LTE atmosphere (see Section \ref{sub:nlte} for more discussion
and examples of this effect). Nevertheless there were almost always
enough useful lines for our method to be successful. Note that we are
not relying on precise knowledge of the star's $U$, $B$, or $V$
magnitude -- which is not always available; \pg\ is one case
-- to calibrate the model flux.

\subsection{Unblended lines}
\label{sub:unblend}

There are various line lists available for the analysis of stellar
spectra; most commonly used is the Kurucz list, although during our
analysis we found that often more accurate oscillator strengths and
wavelengths are available in other databases (e.g.\ \citet{HH95}). In
particular we found that several Co lines had poor term
identifications in the Kurucz list, Fe oscillator strengths were often
out of date, while oscillator strengths for Ti\,\textsc{iii} were
completely inconsistent with our observations.

To determine which lines in our spectra are not blended, we generated
synthetic spectra using only those lines in the Kurucz list that have
been observed experimentally and compared these directly with our
observations. If the continuum was clear on either side of the line
wings, and our line list showed only one possible match, we considered
it to be unblended. Lines that were visibly blended (e.g.\ with overlapping
wings) were ruled out for equivalent width measurement where the blend
could not be be identified.

\subsection{Individual elements}
\label{sub:indiel}

In many regions of our spectra, line blending is extremely severe,
which means that continuum definition is difficult (as discussed
above) and/or that many features are identified with more than one
ion. In this section we discuss the ions we have identified.

\subsubsection{C and N}

\begin{figure}
\vspace{6cm}
\begin{center}
    \includegraphics{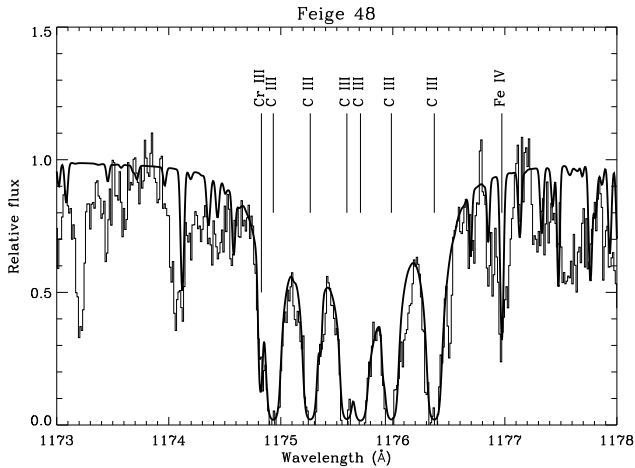}
\end{center}
\caption{Spectrum fit to the C\,\textsc{iii} multiplet at
  1174-1177\,\AA for Feige~48. The model is shown as the thick line. Other
  strong lines due to Cr\,\textsc{iv} and Fe\,\textsc{iv} are marked.}
\label{fig:f48c3}
\end{figure}

\pg\ and \cd\ do not show any carbon lines. In
\cpd\ the C\,\textsc{iii} lines at
1247 and 2297\,\AA\ are present, along with the 1176\,\AA\ multiplet,
while in Feige~48 and Feige~66 these C\,\textsc{iii} lines and the
C\,\textsc{iv} resonance lines around 1550\,\AA\ are seen. Figure
\ref{fig:f48c3} shows the 1176\,\AA\ multiplet of Feige~48 together
with a model spectrum fit; results from spectrum fitting are discussed
further in Section \ref{sec:abu}.

As found in almost all other sdBs, nitrogen lines are strong in our
spectra, although there are fewer lines than seen in optical
spectra. In our two cooler targets, N\,\textsc{iii} lines are visible
at 1183-1184\,\AA\, while the N\,\textsc{iv} line at
1718\,\AA\ is present in Feige~48. In
Feige~66, \pg\ and \cd, lines of N\,\textsc{iii} and
N\,\textsc{iv}, as well as the N\,\textsc{v} resonance lines can be
seen.

\subsubsection{Al and Si}

As has been noted by several authors \citep[e.g.][]{MontrealIII,MontrealV},
the abundance of silicon in sdB stars drops sharply at
$\sim$32\,000\,K. This is also the case for aluminum, although the
available upper limits on abundances are not as strict as for silicon.

Silicon in hot subdwarfs has been discussed in more detail recently by
\citet{SJOT04} and the objects studied here follow previously seen trends. The
detection of the Si\,\textsc{iv} resonance line in \pg\ is discussed
further in Section \ref{sub:comparison}. The Al\,\textsc{iii}
resonance lines are only detectable in Feige~48, although for
\cpd\ the relevant wavelength range is not covered by our spectra.

\begin{figure}
\vspace{6cm}
\begin{center}
    \includegraphics{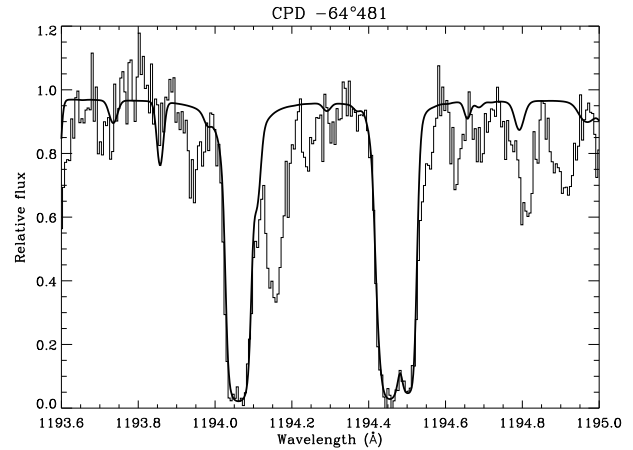}
\end{center}
\caption{The S\,\textsc{iii} resonance lines for \cpd, showing an
  excellent match. The line at $\sim$1194.5\,\AA is one of the Si\,\textsc{ii}
  resonance lines, while the strong line at $\sim$1194.15\,\AA is unidentified.}
\label{fig:SIII}
\end{figure}

\begin{figure}
\vspace{6cm}
\begin{center}
    \includegraphics{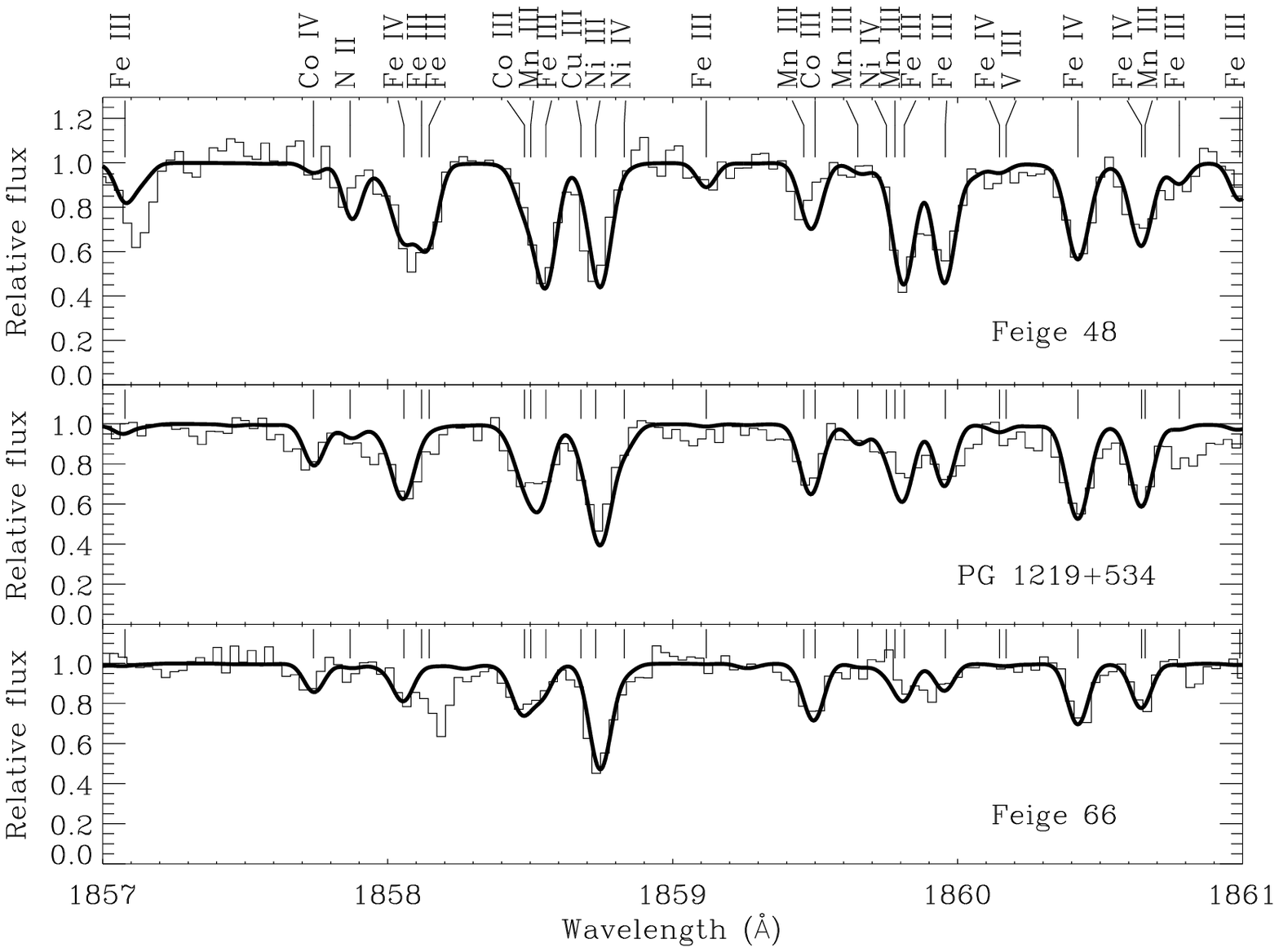}
\end{center}
\caption{A region of spectrum showing blended iron-group lines for
  Feige~48, \pg\ and Feige~66. The line at $\sim$1858.2\,\AA\ is unidentified.}
\label{fig:fe1860}
\end{figure}

\subsubsection{S, Ar, Ca}

The higher resolution of our spectrum of \cpd\ allows us to clearly
distinguish the S\,\textsc{iii} resonance lines (shown in Figure
\ref{fig:SIII}). Because of severe crowding and lower resolution, this
is not possible in the case of Feige~48. The crowding is less severe
for \pg, Feige~66 and \cd\ -- although in the latter the
lines are weak -- so it is possible to make out the S\,\textsc{iii}
lines.

Ar\,\textsc{iii} lines are difficult to measure as they are all quite weak,
and often blended with lines due to iron-group elements. Despite this, we
could measure four lines in Feige~66 and one in \pg; the
other stars have only upper limits.

In the case of calcium, we were able to measure lines of
Ca\,\textsc{iii} for Feige~66, \cd\ and \pg. In
the cooler pair the Ca lines are either very weak or blended, making
measurements difficult.

\subsubsection{The iron group}

Lines of the iron group elements often lie at similar
  wavelengths, making it difficult to measure equivalent widths, and
  necessitating an analysis by spectrum synthesis. An example of this
  can be seen in the 4\,\AA-wide spectrum slices shown in Figure
  \ref{fig:fe1860}.

\textit{Scandium}: There are very few Sc\,\textsc{iii} lines available to
measure in our spectra, making abundances very difficult to measure.
It was only possible to measure the
resonance lines at 1603.064\,\AA\ and 1610.194\,\AA\ for Feige~66 and
\cd. For Feige~48 and \pg\ we could only place upper
limits, and our spectrum of \cpd\ does not cover these
lines.

\textit{Titanium}: The Ti\,\textsc{iii} resonance lines are present in
the spectrum of each star. Some subordinate lines are also
measurable. We note here that the oscillator strengths found in the
Kurucz line list are inconsistent with our observations. Therefore we
have used the values given by \citet{RU97}, which match significantly better.

\textit{Vanadium}: Lines of V\,\textsc{iii} are typically blended or
weak; no lines could be detected at all for Feige~48, allowing only
upper limits to be set.

\begin{figure}
\vspace{12cm}
\begin{center}
    \includegraphics{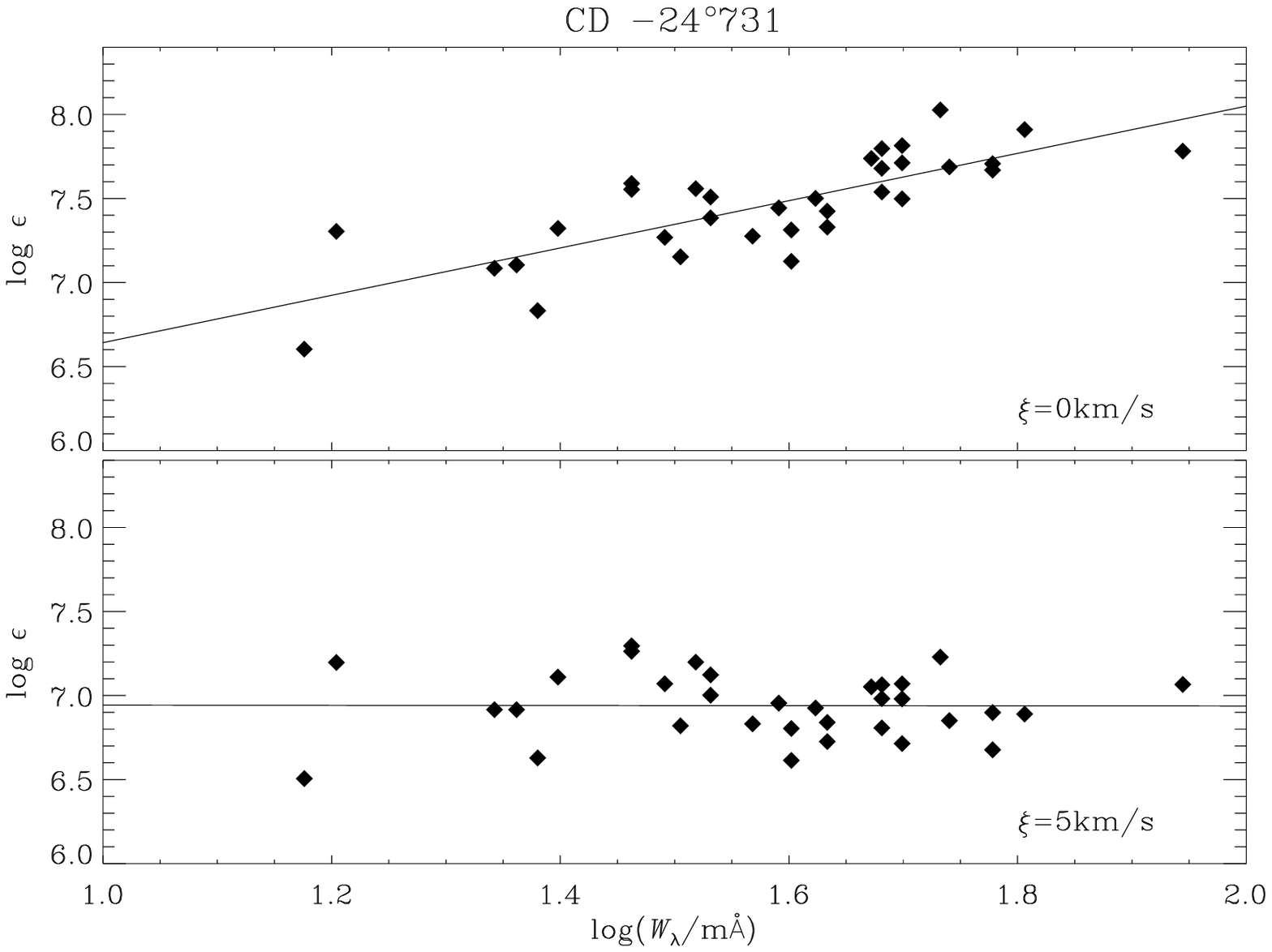}
    \includegraphics{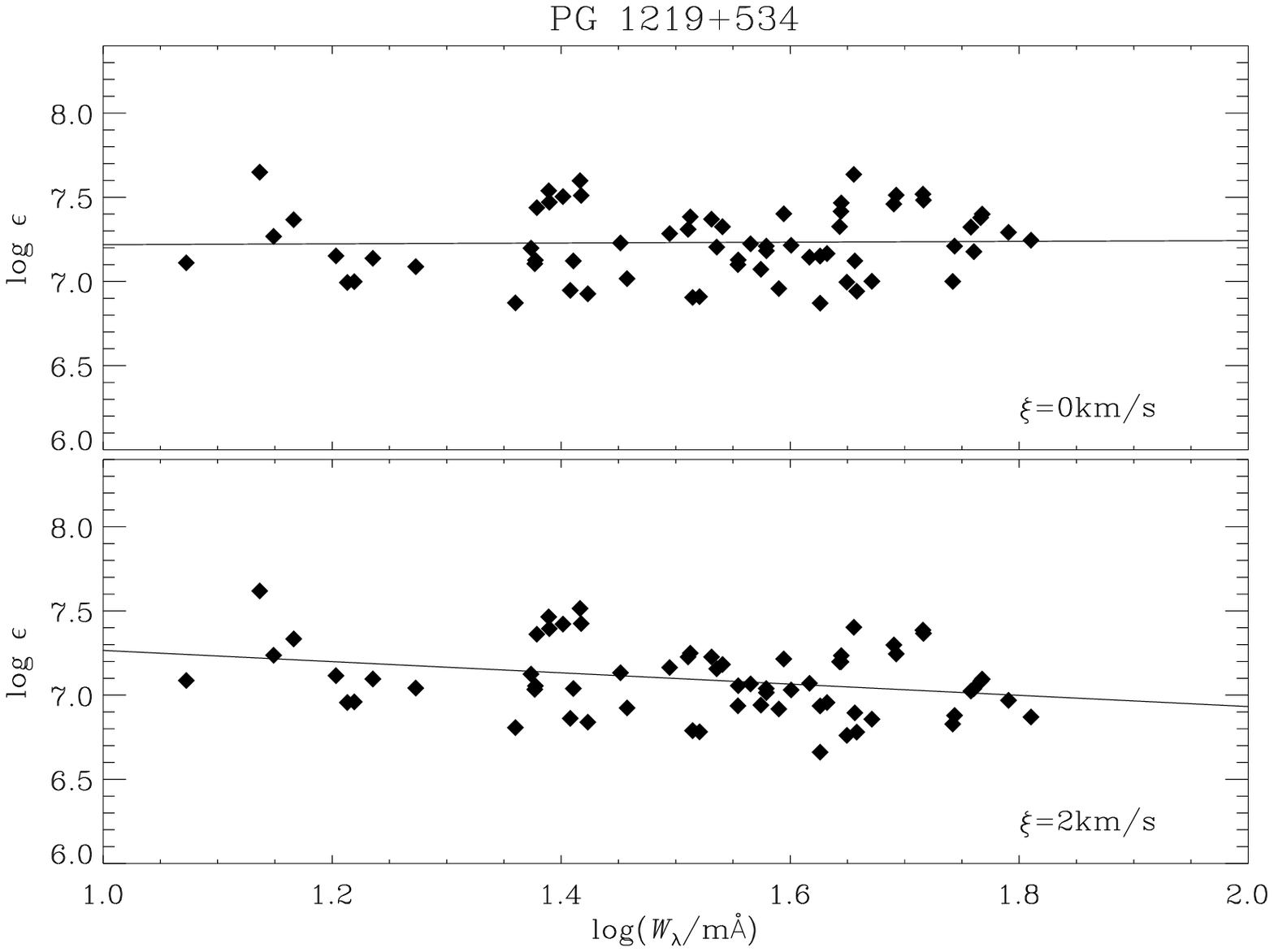}
\end{center}
\caption{Determination of microturbulent velocity of \cd\ and
  \pg using Cr\,\textsc{iii} lines. We find
  $\xi=5^{+2}_{-1}$\,km\,s$^{-1}$ for the former, while the latter is
  consistent with $\xi=0$\,km\,s$^{-1}$.}
\label{fig:vmicro}
\end{figure}

\textit{Chromium}: All of our spectra contain many lines of both
Cr\,\textsc{iii} and Cr\,\textsc{iv} of varying strengths. These
lines, along with those of iron, have been used to determine the
microturbulence discussed in Section \ref{sub:micro}.

\textit{Manganese}: Of all the iron-group elements, manganese causes
the largest difficulties, since it appears that many wavelengths given
in the Kurucz list are not accurate (note that the \citet{HH95} list
uses the same wavelengths). We have examined the list of \citet{UR97},
however they only provide improved oscillator strengths, and also use
the Kurucz wavelengths.

\textit{Iron}: In all spectra there are many Fe\,\textsc{iii} and
Fe\,\textsc{iv} lines measurable. In the two hottest objects in our
sample, Fe\,\textsc{v} lines are also present, although
the accuracy of oscillator strengths is uncertain, since the
abundances derived from some of them differ by several orders of magnitude.

\textit{Cobalt}: Lines of Co\,\textsc{iii} and \textsc{iv} are visible
in our spectra, although often only in crowded regions. It appears
that the hotter objects show more strong lines.

\textit{Nickel}: There are very many nickel lines, particularly of
Ni\,\textsc{iii} and \textsc{iv}. A few Ni\,\textsc{v} lines are also
visible in the spectra of \pg\ and Feige~66, however once
again the accuracy of oscillator strengths is very uncertain.

\textit{Copper}: Cu\,\textsc{iii} and \textsc{iv} are both
present. Neither ion is included in the Kurucz list, so wavelengths
were taken from \citet{HH95}. Oscillator strengths are only available
for Cu\,\textsc{iii}. There are no published oscillator strengths of
Cu\,\textsc{iv}.

\textit{Zinc}: Zn\,\textsc{iii} and \textsc{iv} lines are visible;
however, the latter ion has no published oscillator strengths.

\subsubsection{Gallium, Germanium, Tin and Lead}

The discovery of lines of Ga, Ge, Sn and Pb was made by \cite{SJOT04}. Lines
of Ge\,\textsc{iv} and Sn\,\textsc{iv} have also been seen in hot DA white
dwarfs by Vennes, Chayer \& Dupuis (2005).
Abundances for our four targets are almost all done using spectrum synthesis.
Resonance lines of Ga\,\textsc{iii}, Ge\,\textsc{iv}, Sn\,\textsc{iv}, and
Pb\,\textsc{iv} are present in all spectra. In stars with higher Ga
abundances, Ga\,\textsc{iii} subordinate lines are also present. The
resonance line of Sn\,\textsc{iii} at 1251.387\,\AA\ is strong in
Feige~66, and is present in all stars, but is blended.

One of the resonance lines of Pb\,\textsc{iv} is present in each
object; the other line lies in the wings of Ly$\beta$, which is not
covered by our spectra (but is by \emph{FUSE}).

\begin{table*}
\caption{Abundances for each ion for all five targets.}
\label{tab:4abu}
\begin{center}
\begin{tabular}{l||cr|cr|cr|cr|cr}
 & Feige 66 & & \pg\ & & \cd\ & & Feige 48 & & \cpd
  & \\
\hline
Ion  &  $\log\epsilon$ & $n$ & $\log\epsilon$ & $n$ & $\log\epsilon$ & $n$ &
$\log\epsilon$ & $n$ & $\log\epsilon$ & $n$ \\
\hline
C\,\textsc{iii} &  6.93$\pm$0.28\rlap{$^*$} &   4 & & & & & 7.22$\pm$0.10 & 3 &
  7.54$\pm$0.22 & 2 \\
C\,\textsc{iv}\rlap{$^*$} & 6.5$\pm$0.5 & 2 & $<$3.1 & & $<$2.7 & &
6.9$\pm$0.5 & & &  \\
N\,\textsc{ii} & & & & & & & & & 7.87$\pm$0.14 & 3 \\
N\,\textsc{iii} &  7.69$\pm$0.23 &   2 & 7.64$\pm$0.42 &  4 & 7.60$\pm$0.47 &
 6 & 7.62$\pm$0.38 & 3 & 7.40$\pm$0.00 & 2 \\
N\,\textsc{iv} & 7.65$\pm$0.00 & 1 & & & & & & \\
N\,\textsc{v}   & 7.60$\pm$0.11 & 2 & 7.58$\pm$0.21 &  2 & 8.26$\pm$0.15 & 2
 & & \\
Al\,\textsc{iii} & $<$3.5 & & $<$3.8 & & $<$3.6 & & 5.51$\pm$0.10 & 3 \\
Si\,\textsc{iii} & & & & & & & 6.43$\pm$0.31 & 8 & 6.38$\pm$0.37 & 3 \\
Si\,\textsc{iv}\rlap{$^*$}  & $<$2.0 & & 3.63$\pm$0.17 &  2 & $<$2.0 & & 6.16$\pm$0.25 &
 3 \\
P\,\textsc{iii} & & & & & & & 4.41$\pm$0.00 & 1 & 4.81$\pm$0.09 & 3 \\
P\,\textsc{iv} & 3.95$\pm$0.00 & 1 & & & & & & \\
S\,\textsc{iii} & 7.69$\pm$0.46 & 4 & 7.00$\pm$0.11 &  3 &
5.7$\pm$0.00 & 1 & & & 6.29$\pm$0.27 & 3 \\
Ar\,\textsc{iii} &  7.86$\pm$0.24 &   4 &  7.02$\pm$0.00 &  1 & $<$7.0
& & $<$7.0 & & & \\
Ca\,\textsc{iii} &  8.09$\pm$0.20 &  20 &  7.73$\pm$0.23 & 10 & 7.50$\pm$0.22
& 8 & 7.31$\pm$0.00 & 1 & 6.97$\pm$0.00 & 1 \\
Sc\,\textsc{iii} & 5.17$\pm$0.51 & 2 & $<$5.0 & & 4.61$\pm$0.00 & 1 & $<$3.0 & & & \\
Ti\,\textsc{iii} &  6.98$\pm$0.22 &   10   &    6.51$\pm$0.20 &  2 &
5.77$\pm$0.06 & 3 & 5.16$\pm$0.20 & 4 & 5.68$\pm$0.37 & 6 \\
Ti\,\textsc{iv}  &  6.91$\pm$0.18 &   5   &    6.73$\pm$0.00 &  1 &
5.87$\pm$0.15 & 3 & 5.21$\pm$0.18 & 2 & 5.75$\pm$0.07 & 2 \\
V\,\textsc{iii} &  6.42$\pm$0.22 &  11   &    5.80$\pm$0.00 &  1  & &
& & & 4.48$\pm$0.15 & 4 \\
V\,\textsc{iv}  &  6.29$\pm$0.14 &   7   &    6.37$\pm$0.34 &  2 &
5.77$\pm$0.37 & 9 & & & & \\
Cr\,\textsc{iii} &  7.29$\pm$0.19 &  65   &    7.23$\pm$0.17 & 44 &
6.96$\pm$0.23 & 44 & 5.90$\pm$0.18 & 29 & 6.20$\pm$0.30 & 33 \\
Cr\,\textsc{iv}  &  7.31$\pm$0.23 &  25	&   7.46$\pm$0.14  & 17 &
7.24$\pm$0.47 & 28 & 6.41$\pm$0.28 & 7 & 6.54$\pm$0.20 & 10 \\
Mn\,\textsc{iii} &  6.02$\pm$0.19 &  12	&   7.25$\pm$0.22 & 46  &
6.89$\pm$0.36  & 19 & 5.62$\pm$0.21 & 13 & 5.54$\pm$0.21 & 15 \\
Mn\,\textsc{iv}  &  5.99$\pm$0.51 &   2	&   7.72$\pm$0.11 &  3  &
7.12$\pm$0.54  & 7 & & & 5.86$\pm$0.18 & 2 \\
Fe\,\textsc{iii} &  6.46$\pm$0.17 &  26	&   7.16$\pm$0.22 & 69  &
6.66$\pm$0.26  & 23 & 7.68$\pm$0.19 & 107 & 7.38$\pm$0.33 & 20 \\
Fe\,\textsc{iv}  &  6.40$\pm$0.10 &  2	&   7.43$\pm$0.22 & 23  &
7.03$\pm$0.25  & 23 & 7.73$\pm$0.21 & 10 & 7.52$\pm$0.23 & 11 \\
Fe\,\textsc{v}   & & & 7.38$\pm$0.41 &  2  & &  & & & & \\
Co\,\textsc{iii} &  6.46$\pm$0.33 &  39 &  6.73$\pm$0.20 & 20  &
6.13$\pm$0.39  & 20 & 5.73$\pm$0.25 & 5 & & \\
Co\,\textsc{iv}  &  6.32$\pm$0.22 &   20 &  6.72$\pm$0.13 & 7  &
6.28$\pm$0.21  & 11 & 6.05$\pm$0.25 & 3 & & \\
Ni\,\textsc{iii} &  6.86$\pm$0.15 &  32 &  7.39$\pm$0.18 & 32  &
6.20$\pm$0.48  & 12 & 6.63$\pm$0.22 & 28 & 6.39$\pm$0.25 & 6 \\
Ni\,\textsc{iv}  & 6.81$\pm$0.25 & 32 &  7.41$\pm$0.27 & 19  & 6.28$\pm$0.28 &
14 & 6.75$\pm$0.21 & 7 & 6.50$\pm$0.26 & 9 \\
Ni\,\textsc{v}   & 6.91$\pm$0.55 & 10 &  7.60$\pm$0.14 & 5  & & & & & & \\
Cu\,\textsc{iii} & 6.27$\pm$0.38 & 8 & 6.36$\pm$0.36 & 5 & & &5.25$\pm$0.40 &
3 & & \\
Zn\,\textsc{iii} &  6.64$\pm$0.35 &   15 &  6.68$\pm$0.36 &  5  &
5.97$\pm$0.50 & 2 & 5.30$\pm$0.22 & 6 & 5.33$\pm$0.34 & 6 \\
Ga\,\textsc{iii} & 5.53$\pm$0.06 & 2 & 5.75$\pm$0.04 & 2 & 5.23$\pm$0.21 & 2 &
4.70$\pm$0.00 & 1 & 3.63$\pm$0.00 & 1 \\
Ge\,\textsc{iv}  & 5.21$\pm$0.05 & 2 & 4.99$\pm$0.03 & 2 & 4.98$\pm$0.01 & 2 &
4.02$\pm$0.08 & 2 & 3.57$\pm$0.10 & 2 \\
Sn\,\textsc{iv}  & 4.12$\pm$0.00 & 1 & 4.08$\pm$0.00 & 1 & 3.11$\pm$0.00 & 1 &
2.94$\pm$0.00 & 1 & 2.86$\pm$0.00 & 1 \\
Pb\,\textsc{iv}  & 4.7$\pm$0.00 & 1 & 4.6$\pm$0.00 & 1 & 4.3$\pm$0.00 & 1 &
3.8$\pm$0.00 & 1 & 3.43$\pm$0.00 & 1 \\
\hline
\end{tabular}
\end{center}
\hspace{2cm}
$^*$Based on non-LTE-affected resonance lines; see Section
\ref{sub:nlte}.
\end{table*}

\section{Abundance analysis}
\label{sec:abu}

\begin{figure}
\vspace{6cm}
\begin{center}
    \includegraphics{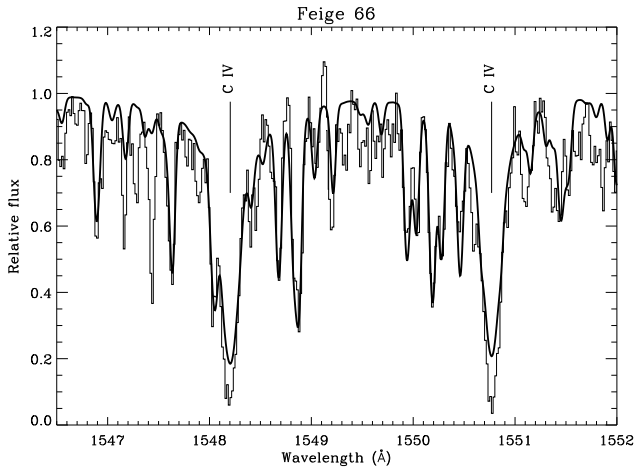}
\end{center}
\caption{Spectrum fit to the C\,\textsc{iv} resonance lines at 1550\,\AA\ for
  Feige~66. The model is shown as the thick line. The line cores of the model
  do not match the observations, and the abundance is set around 0.8 dex lower
  than that derived from C\,\textsc{iii} lines. See text for details.}
\label{fig:f66c4}
\end{figure}

\begin{figure*}
\vspace{22cm}
\begin{center}
    \includegraphics{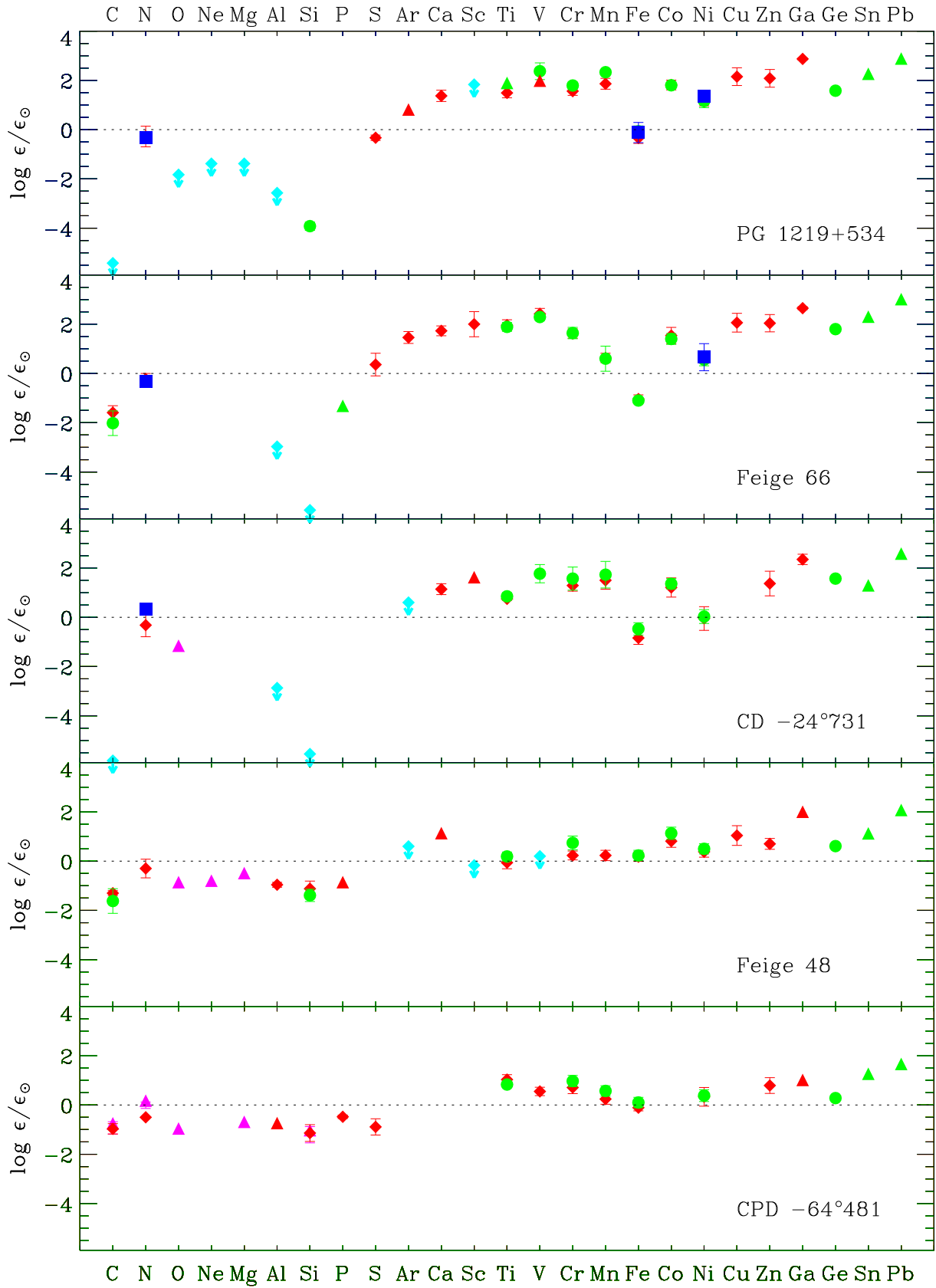}
\end{center}
\caption{Abundances measured for our five targets. Magenta symbols
  represent values determined using singly ionised lines, green
  represents doubly ionised lines, red
  represents triply ionised lines, blue denotes quadruply ionised
  lines and cyan represents upper limits. Note the generally excellent
agreement between different ionisation stages.}
\label{fig:4abu}
\end{figure*}

In this section we comment on results and trends for individual or groups of
elements. The abundances are given in Table \ref{tab:4abu} and plotted
relative to the sun in Figure \ref{fig:4abu}.  The errors
  given were determined using a simple mean and standard deviation
  based on the individual measurement for each line. An error of zero
  represents a measurement taken from only one line. The solar values
were taken from \citet{GS98}. The O, Ne and Mg abundances of the
programme stars are taken from HRW for \pg\ and Feige 48, while for
\cd\ and \cpd, they are from Edelmann (private communication), along
with the Al abundance for the latter star.

Before discussing our derived abundances in more detail, however, it
is necessary to first consider non-LTE effects and the effect of
microturbulence on the absorption lines in our spectra.

\subsection{NLTE effects}
\label{sub:nlte}

Most of the spectral lines we are dealing with in this paper are subordinate,
i.e. they are between excited levels and do not involve the ground 
level. The
resonance lines (transitions from the ground level) of some ions are present
however, and it is these lines that are most sensitive to departures from our
assumption of local thermodynamic equilibrium (LTE). Ions of
iron-group elements do not appear to be affected, but
species such as C\,\textsc{iv} and Si\,\textsc{iv}, there are noticeable
effects. In Figure \ref{fig:f66c4} we show the C\,\textsc{iv}
resonance lines at 1548\,\AA\ and 1550\,\AA\ of Feige~66 as an
example. In this star and Feige~48 the line cores of these lines are
matched by the abundances derived from the C\,\textsc{iii} lines,
however the wings are much too broad. Reducing the C abundance results
in an improved fit in the wings but a core that is too shallow. This
is most likely a non-LTE effect. Detailed model atom calculations are
needed to confirm this, but are beyond the scope of this paper. There
may also be an interstellar component present, at least in Feige~66,
which has a radial velocity of only $\sim -6$\,km\,s$^{-1}$; however,
this is not the case for Feige~48, and the effect is still present.

In the case of C\,\textsc{iii} all three lines/multiplets can be
matched well with a model spectrum at the listed abundance for
Feige~48, however for Feige~66 things are not so simple. The
1174-1177\,\AA\ multiplet is apparently matched well; the
1247\,\AA\ line is not, while the 2297\,\AA\ line matches in the wings
but not in the line core. Because Feige~66 is around 5000\,K hotter
than Feige~48, we again attribute these problems to non-LTE effects.

The iron-group lines we have used for our abundance analysis are not
resonance lines (with the exception of Ti\,\textsc{iii} where some
resonance lines were used), so are
not likely to be strongly affected by non-LTE effects.

\subsection{Microturbulence}
\label{sub:micro}

Since we have measured the equivalent widths of many lines for several ions,
we can investigate the microturbulence $\xi$ for each star. Both of the
pulsating sdBs, along with \cpd, are consistent with $\xi=0$\,km\,s$^{-1}$;
however, the two other non-pulsators both have $\xi>0$\,km\,s$^{-1}$. The top
panels of Figure \ref{fig:vmicro} shows the difference between
$\xi=0$\,km\,s$^{-1}$ and the value we derive $\xi=5^{+2}_{-1}$\,km\,s$^{-1}$
for \cd; the bottom panels show the effect of non-zero microturbulence for
\pg. For Feige~66 we find $\xi=2\pm1$\,km\,s$^{-1}$. These values have been
derived using both Cr\,\textsc{iii} and Fe\,\textsc{iii} lines.

\subsection{Abundances}
\label{sub:comparison}

Before looking at the differences in iron-group abundances of our
targets, we first search for and examine trends amongst the lighter
elements.

\subsubsection{Light metals} 

As has been found by many previous studies, carbon abundances range
from virtually none at all to slightly below the solar value. For
\pg\ and \cd\ the absence of the C\,\textsc{iv} resonance
lines indicates carbon depletion by 10$^5$ or more. Because we cannot
at the moment account for NLTE effects (see \ref{sub:nlte} above)
these have to be regarded as order of magnitude estimates. We note
here that simple NLTE calculations were done by \citet{HHH84} for the
sdO star Feige~110 ($T_{\mathrm{eff}}=40\,000$\,K) and a
similar upper limit was derived.

In the case of nitrogen, the abundances are slightly below the solar
value; this is also in keeping with previous analyses of sdB optical
spectra.  We note in passing that we could not easily fit the
N\,\textsc{iv} line at 1718.551\,\AA\ in all of our sample because of
severe crowding -- only in Feige~66 is the line isolated -- however
the strength of the line is approximately consistent with the
abundances derived from the other N ionisation stages. Note that the
nitrogen abundances from three different ionisation stages are fully
consistent for Feige~66.  

For Feige~48 it was not possible to measure the sulfur abundance because the
resonance lines are blended with a large number of metal lines that are not in
the Kurucz list of observed lines or the Hirata \& Horaguchi list. This effect
is seen to a lesser extent in \pg, making a spectrum fit possible,
and hardly seen at all in Feige~66. Because of the absence of heavy crowding
in the latter star, we suggest these may be iron lines. We examined the
effect of including the ``computed'' (but not observed) lines in the Kurucz
list: the match is better for Feige~48, but it is still not possible to
measure abundances using the S\,\textsc{iii} lines.

\subsubsection{Iron group}

As can be seen from Table \ref{tab:4abu}, plenty of lines of doubly
ionised atoms (in particular Cr\,\textsc{iii}, Fe\,\textsc{iii} and
Ni\,\textsc{iii}) have been used for the abundances analysis, for Feige~48
more than 100 Fe\,\textsc{iii} lines were utilised. Moreover, triply
ionised atoms can be used to determine abundances for all iron group
elements as well (except for V\,\textsc{iv} and Mn\,\textsc{iv} in
Feige~48 and V\,\textsc{iv}, and Co\,\textsc{iv} in \cpd), which in
some stars are also quite numerous (e.g. 32 Ni\,\textsc{iv} lines in
Feige 66). Even four times ionised atoms have been used for analysis
of nickel in Feige~66 and \pg\ and of iron in \pg. It is worthwhile
noting that these ionisation equilibria are very well matched,
e.g. the three ionisation stages of Ni  agree to within 0.1 dex for
Feige 66 and to within 0.2 dex for \pg. For other ions of the iron
group, we find:

\begin{itemize}
\item Fe\,\textsc{iii} and Fe\,\textsc{iv} as well as Ni\,\textsc{iii} and
  Ni\,\textsc{iv}, respectively, agree very well (0.1~dex, typically) for
  Feige~48 and \cpd\ and to within error limits in \cd\ (0.37~dex for iron).

\item Cr\,\textsc{iii} and Cr\,\textsc{iv} agree to better than 0.3~dex except
  for Feige~48 (0.5~dex).

\item Mn\,\textsc{iii} and Mn\,\textsc{iv} are in perfect agreement for Feige
  66, differ by about 0.3~dex in \cd\ and \cpd. Only for \pg\ they deviate
  beyond the adopted error ranges. 

\item Co\,\textsc{iii} and Co\,\textsc{iv} also match very well, to better
  than 0.2~dex, except for Feige~48 which has much less Co lines than the
  other programme stars.

\item Although only few titanium lines could be used for the analysis, the
  results from Ti\,\textsc{iii} and Ti\,\textsc{iv} match extremely
  well for all programme stars.
\end{itemize}

We regard the very good match of many ionisation equilibria as evidence that
systematic errors of the metal abundances and of the effective 
temperature are small. This can also be seen in the good match between
synthetic spectrum and observations for three objects shown in Figure
\ref{fig:fe1860}.

While iron is found to be nearly solar (\pg, Feige~48, \cpd),
slightly depleted in \cd\ and subsolar in Feige~66 by a factor of ten, all
other elements of the iron group are enhanced by between 0.5 and 2.5 dex with
respect to solar values. The enhancements are large in Feige~66 and \pg, but
mild for the three others.

\begin{figure}
\vspace{6cm}
\begin{center}
    \includegraphics{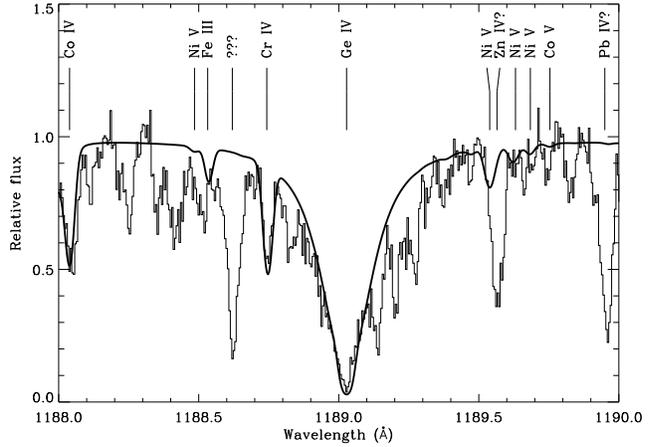}
\end{center}
\caption{Spectrum fit to the Ge\,\textsc{iv} resonance lines at 1189\,\AA\ for
  Feige~66. The model is shown as the thick line. }
\label{fig:f66ge4}
\end{figure}

\subsubsection{Gallium, Germanium, Tin and Lead} 

The heavy metals Ga, Ge, Sn and Pb are all enriched with
respect to the Sun in all stars, reaching as high as 2.9~dex for Ga in 
\pg\ or 2.75~dex for Pb in Feige~66. An example of a fit of the
Ge\,\textsc{IV} 1189\,\AA\ line is shown in Figure
\ref{fig:f66ge4}. There are several strong lines also present that are
either not identified or have no atomic data. Possible identifications
are shown in the figure.

\subsection{Comparison with previous work}
\label{sub:previous}

As mentioned earlier, many of the objects we have observed for this project
have been studied with optical spectra before -- \pg\ and
  Feige~48 \citep{HRW00}, and \cpd\ and \cd\ (Edelmann, private
  communication) -- or with a noisy \emph{IUE} spectrum and poor atomic data
-- Feige~66 \citep{BHS82}. It is useful here to compare our results with these
previous studies, as well as with any general trends seen amongst sdBs as a
group.

Firstly, for the two pulsators \pg\ and Feige~48, our results compare
very well with those of HRW for the limited cross-over that exists
between elements. All of HRW's abundances are, within errors,
consistent with ours, and because we could use resonance lines of
several ions, we have been able to place stronger upper limits on C, Al
and Si for \pg. In the case of
Feige~48 the abundances we derive agree well with those measured by
HRW, and are within $\pm0.04$\,dex.

The only element we have had difficulty
deriving an abundance for is sulfur, for the reasons discussed earlier.

In the case of Feige~66, \citet{BHS82} used \emph{IUE} spectra with
lower resolution than ours, and obtained somewhat different results to
ours. It is difficult to compare the two sets of abundances since
\citeauthor{BHS82} do not give error estimates explicitly; however, if
we assume the same errors as those given for another sdB, HD\,149382,
in the same paper, our values are not so discrepant after all. At the
time of \citeauthor{BHS82} study, the quality of atomic data was lower
than it is at present (although in many cases it is still insufficient
or of low accuracy), so their errors are very large, particularly for
the iron group.

Recently, \citet{EHN06} has carried out an analysis
of \cpd\ and \cd\ based on high-resolution optical echelle
spectra. For both stars their abundances are consistent with
those determined here.

Finally, we can also compare any apparent trends seen here with the work of
\citet{HE04}, especially for the lighter elements plus iron.
Both our results here and the work of \citet{HE04} suggests that aluminum
follows the same trend as silicon.

\subsection{Heavy element abundance trends?}
\label{sub:fetrend}

\begin{figure}
\vspace{7.5cm}
\begin{center}
    \includegraphics{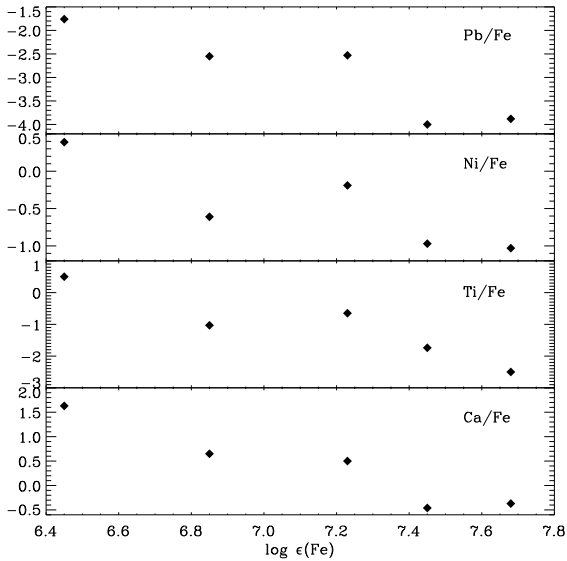}
\end{center}
\caption{Trends in heavy element abundances with iron abundance.}
\label{fig:fetrend}
\end{figure}

An interesting result not immediately obvious from the abundances in Table
\ref{tab:4abu} is a possible anti-correlation between heavy element abundance
and iron abundance. The spread in iron abundance is 1.25 dex from the most
iron-poor (Feige~66) to the most iron-rich star (Feige~48), while the variation
(X/Fe) is between 2 and 3 dex (except for Mn and Ni). Therefore this trend
does not simply reflect the trend in iron abundance. 
Four examples of this effect are
shown in Figure \ref{fig:fetrend} for Ca/Fe, Ti/Fe, Ni/Fe and Pb/Fe while
Table \ref{tab:fetrend} shows data for all elements; where there are two or
more ionisation stages, a weighted average is presented. The same basic trend
is seen for all elements in the iron-group, except for manganese and perhaps
nickel, and it is also seen for the heavier elements gallium, germanium, tin
and lead. Caution must
be applied here however, since we have only five objects in our sample. A more
detailed analysis may be possible for at least titanium, since several sdB
stars show Ti\,\textsc{iii} lines in their optical spectra (H. Edelmann,
C. Karl, private communication). If these trends are real we believe they are
caused by a combination of radiative levitation, gravitational settling and a
weak stellar wind, and therefore would help constrain diffusion models. We
urge more theoretical work in this area.

\begin{table}
\caption{Average iron abundance and ratio of heavy element abundances to iron
  for the five targets stars. The columns are as follows: F66 = Feige~66; CD =
  \cd; PG = \pg; CPD = \cpd; F48 = Feige~48. Abbreviations: n-v = non-variable,
  var. = variable.}
\label{tab:fetrend}
\begin{center}
\begin{tabular}{l|rrr|rr}
 & F66 & CD & PG & CPD & F48 \\
 & n-v & n-v & var. & n-v & var\\
\hline
Fe & 6.46 & 6.85 & 7.23 & 7.43 & 7.68 \\
\hline
Ca/Fe & 1.63    & 0.65    & 0.50     &  $-$0.46  &  $-$0.37 \\
Ti/Fe & 0.50    & $-$1.03 & $-$0.65  &  $-$1.74  &  $-$2.50\\
V/Fe  & $-$0.09 & $-$1.08 & $-$1.08  &  $-$2.95  &   ---   \\ 
Cr/Fe & 0.84    & 0.23    & 0.06     &  $-$1.15  &  $-$1.68 \\
Mn/Fe & $-$0.44 & 0.10    & 0.05     &  $-$1.85  & $-$2.06 \\
Co/Fe & $-$0.05 & $-$0.69 & $-$0.50  &    ---    & $-$1.83 \\
Ni/Fe & 0.39    & $-$0.61 & $-$0.19  &  $-$0.97  & $-$1.03 \\
Cu/Fe & $-$0.19 &  ---    & $-$0.87  &    ---    & $-$2.43\\
Zn/Fe & 0.18    & $-$0.88 & $-$0.59  &  $-$2.10  & $-$2.38\\
Ga/Fe & $-$0.93 & $-$1.62 & $-$1.70  &  $-$3.80  & $-$2.98 \\
Ge/Fe & $-$1.25 & $-$1.87 & $-$2.02  &  $-$3.86  & $-$3.66\\
Sn/Fe & $-$2.34 & $-$3.74 & $-$3.11  &  $-$4.57  & $-$4.74 \\
Pb/Fe & $-$1.76 & $-$2.55 & $-$2.53  &  $-$4.00  & $-$3.88 \\
\hline
\end{tabular}
\end{center}
\end{table}

\subsection{Comparison of abundances of pulsator/non-pulsator pairs}
\label{sub:pnpcomp}

\begin{figure}
\vspace{11.5cm}
\begin{center}
    \includegraphics{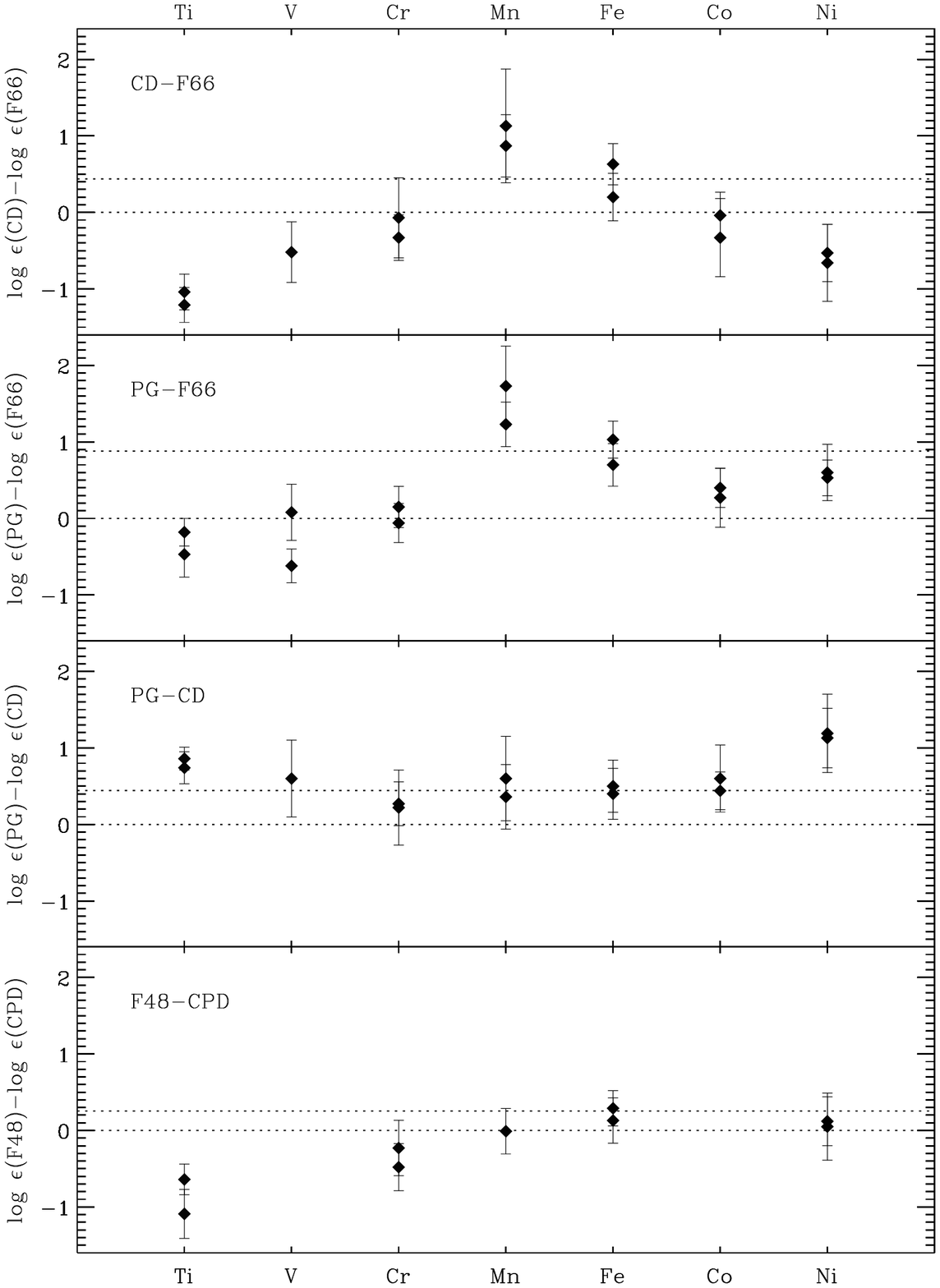}
\end{center}
\caption{Comparison of iron-group abundances for our
  pulsator/non-pulsator pairs (F48\,=\,Feige~48; CPD\,=\,\cpd; 
  PG\,=\,\pg;
  CD\,=\,\cd; F66\,=\,Feige~66). The dotted lines denote
  equal abundances and the difference in iron abundance. The
  differences between  \pg, \cd\ and Feige~66 are quite large
  -- particularly for Fe, Ni and Mn -- while the pairs 
  \pg\ $\&$ \cd\ and Feige~48 $\&$ \cpd\ show
  quite similar abundances, in particular when the differences in iron abundance
  are accounted for (see text).}
\label{fig:ironcomp}
\end{figure}

 The results of our spectral analysis  
allow us to compare the abundance
patterns of four pairs of stars, having similar effective temperature:
(i) the ``cool'' pulsator/non-pulsator pair Feige~48/\cpd, 
(ii) the ``hot'' pulsator/non-pulsator 
pair \pg/\cd, (iii) the ``hot'' pulsator/non-pulsator pair \pg/Feige~66 and 
(iv) the pair of
two ``hot'' non-pulsators Feige~66/\cd.
 We plot the relative iron group abundances for all four combinations 
in Fig.~\ref{fig:ironcomp}).

(i) If we compare Feige~48
and \cpd\ (the bottom panel of Figure \ref{fig:ironcomp}), we
find basically no difference. Not only are the abundances of 
the iron group elements similar, but the lighter and
heavier elements also agree reasonably well.
The only exception is gallium which is ten times higher in Feige~48 than in
\cpd.  Also looking at abundance ratios of iron group elements 
with respect to iron (X/Fe, Table \ref{tab:fetrend}) reveals, that the 
patterns for Feige~48 and \cpd\ match each other reasonably well. 
This would indicate that there
is no difference in iron group abundances of a pulsator and a non-pulsator.

(ii) The comparison of ``hot'' pulsator \pg\ to the non-pulsator \cd\ reveals 
large differences for some light elements, silicon and sulphur being much
less in \cd\ (more than 1.6 dex and 1.3 dex, respectively). Significant
differences are also detected for the iron group as well as for the
heavy elements. They, however, vanish almost if the difference in iron
content is accounted for (except for Ni/Fe and Sn/Fe, see
Table~\ref{tab:fetrend}). Since the iron group elements 
are probably the drivers for pulsators, the comparison would indicate as in 
the previous case that there is little difference (if at all) between
a pulsator and a non-pulsator.

(iii) The comparison of the ``hot'' pulsator \pg\ to the non-pulsator Feige~66
yields a completely different picture. Abundances of several iron group elements
differ (see Table~\ref{tab:4abu}). Even if we take into account the large 
difference in iron abundance (factor of 7) between the two stars, the 
abundance patterns remain dissimilar (see Table~\ref{tab:fetrend}). Amongst the 
light elements there is a huge difference in carbon
abundance. \pg\ has less than 5000 times as much C as Feige~66
has. The abundances of the heavy elements (Ga, Ge,
Sn, and Pb), however, are quite similar. In contrast to the previous comparisons
in (i) and (ii) this would imply that indeed pulsators and non-pulsators have 
\emph{different} iron group abundances.     

(iv) The comparison of the two ``hot'' non-pulsators Feige~66 and \cd\ also 
reveals large differences in the abundances of iron group elements which 
persist if we account for the different iron abundance 
(see Table~\ref{tab:fetrend}). Amongst the light elements as well as among the 
heavy elements there are large differences as well, which persist if we account for the 
different iron abundances.

The last finding is discouraging. If the difference among non-variable 
stars of similar $T_{\mathrm{eff}}$ and  $\log g$ are as large as we observe, 
the comparison between a pulsator and a non-pulsator is rendered arbitrary. 

The theory of \citet{CFB97a} does not specifically rule out iron-group
elements other than iron itself as responsible for pulsation
driving. In particular Ni must be considered as it has substantially
more UV and FUV lines than iron. Indeed, in the case of \pg\ the nickel 
abundance is high
enough that it could make a significant contribution to the opacity,
at least as much as iron. When we consider Feige~48 and \cpd, we find
that perhaps the difference in the \emph{combined} opacity of iron and
nickel between the two stars is significant enough to discriminate
between pulsator and non-pulsator. This leads us to ask: could we
theoretically have a pulsator with lower iron, but much higher
nickel or other iron-group abundance than a non-pulsator?

\section{Further discussion}
\label{sec:disc}

As seen in Section \ref{sub:tempfix}, enhanced metal abundances can
have an effect on $T_{\mathrm{eff}}$ and $\log g$ determination,
especially for stars showing both neutral and singly ionised
helium. It is also clear that simply scaling models from solar
metalicity ODFs is insufficient; opacity sampling is required for more
accurate measurements, since while Fe abundances are approximately
solar, elements such as Ni and Mn -- which have a significant opacity
contribution -- are almost always enhanced. A preliminary study has
been done in this direction by \citet{BJ06} and \citet{PNE06}.

The effect of these abundance patterns may be apparent for atmosphere
models, but what about stellar evolution and pulsation models? We urge
evolution theorists to investigate the effect of non-solar opacity
distributions on hot subdwarf evolution. In a similar vein, interior models
that include differential internal rotation and/or magnetic fields should also
investigate their effect on diffusion.

In a study of elements beyond the iron group in hot subdwarfs,
\citet{SJOT04} proposed a solution to  the silicon problem discussed in
Section 4.3.2. Triply ionised elements of the same group in the
Periodic Table as silicon -- germanium, tin and lead -- are present in
almost all sdB spectra at \emph{all} temperatures. Si\,\textsc{iv} on the
other hand, almost completely disappears above 32\,000\,K. This shows
that arguments where silicon is ionised to  nobe gas configuration, 
i.e. to Si\,\textsc{v}, and then
sinks deeper into the atmosphere, cannot be correct. If this were the
case, these heavier elements, which should feel the same  low radiative
forces as silicon, should also sink. Instead, \citet{SJOT04} suggested
that above $\sim$32\,000\,K silicon could be carried away by a weak
fractionated stellar wind, whereas the heavier elements should stay
behind. Indeed, \citet{Unglaub06} has found that a one-component,
uniform wind is inconsistent with observations, and that multicomponent
calculations are required, although this depends on
  surface gravity and mass-loss rate. When we examine the abundances
of Ge, Sn and Pb, we find that they are all higher in the hot stars
(Feige~66, \cd, and \pg) that show little or no silicon, than in the
cooler stars (Feige~48, \cpd) that do. This is what one would
qualitatively expect from the hypothesis of \citet{SJOT04}. Using
\emph{FUSE} spectra, \citet{CFF06} found many other sdBs with similar
properties, but also some exceptions. The case of C\,\textsc{iv} is
not at all straightforward, since it is very difficult to explain two
spectroscopically similar stars, one with measureable silicon but no
carbon (\pg), and the other with measureable carbon but no silicon
(Feige~66). \cd\ has no carbon and no silicon, compounding the
problem. The reason for this remains a mystery, and presents
  a challenge to the fractionated wind hypothesis, especially since
  carbon is also in the same group as silicon, germanium, tin and
  lead, and should in principle feel the same radiative forces.

It is also worth noting that \pg, one of our pulsators, shows measurable
traces of silicon, despite being hotter than 32\,000\,K, while the
spectroscopically similar  stars Feige 66 and \cd\ do not. Could
this be an indicator of the age of the sdB? If the star has a weak
fractionated wind, as suggested by \citet{SJOT04}, then perhaps one of
the differences between \pg\ and Feige~66/\cd\ is age. In order to
determine this, however, diffusion models including silicon are required.

It seems clear that radiative acceleration is bringing heavy elements to the
surface in these sdBs, and perhaps even expelling them through a stellar
wind. We can at least qualitatively understand this in the case of the
iron-group and heavier elements: in the hotter stars heavy elements are much
more enhanced than in the cooler stars. We must also consider, however, that
two of our ``hot'' stars are apparently single, while both of our ``cool'' 
stars are in close binary systems. Does this have any effect on the 
abundance patterns?
It is unclear what the cause  
for the differences  among the ``hot'' stars, Feige~66, \cd, and \pg, is, 
but binarity is one possibility. 
Our small
number statistic  is the main limiting factor for this discussion, so we must
look forward to the abundance analyses of Edelmann et al. (2006), who
have studied a larger number of binaries and single stars,
albeit with optical spectra, which restricts the available heavier
elements to iron and sometimes titanium.

\section{Summary and Conclusions}
\label{sec:conc}

We have analysed high-resolution UV echelle spectra of five hot
subdwarf B stars, two of which are member of the short-period, pulsating
V361\,Hya class. 
 Abundances of no less than 25 elements including the iron group and 
  even 
  heavier elements such as tin and lead have been determined using 
  LTE curve-of-growth and spectrum 
  synthesis techniques. 
  Our investigation was initiated to test the hypothesis that a
correlation exists between the abundances of iron-group elements in
sdB stars and pulsation. We have compared a hot pulsator (\pg) with a
 two non-pulsators with similar stellar parameters (Feige~66 and \cd) 
and a cooler pulsator
(Feige~48) with a similar non-pulsator (\cpd), and found no consistent
differences between the members of each pair. 
 The heavy element abundance pattern of \cd\ comes close to that
observed for \pg\ except for its low iron and nickel. Feige~66 has an even lower
iron abundance, but its heavy metal abundance pattern does not match that of
\pg\ at all. In other words the abundance patterns of two non-variable stars 
of similar temperature and gravity are too dissimilar for a conclusive
comparison with a pulsator. This result  
leads us to suspect
that there must be another, as yet unknown, discriminating factor between
pulsating and non-pulsating sdB stars. 

More generally, we have uncovered a potential solution to the discrepant
effective temperatures from Balmer lines and helium ionisation equilibria,
with a significant improvement found using supersolar metallicity
models. Opacity sampling in place of distribution functions will sure lead to
all model line profiles matching observations. Additionally, our spectra show
light element abundance patterns typical for sdBs: carbon varies from
virtually none to around 1 dex below solar, while nitrogen is within 0.5 dex
of the solar value. There is evidence to support the fractionated weak stellar
wind hypothesis of \citet{SJOT04}, as the heavy element abundances increase
with temperature. We also find an interesting anti-correlation between the
abundance of iron and the heavy element abundance relative to iron. More
observations are needed to confirm this trend.

So somewhat frustratingly we must conclude that we cannot provide any
insights into pulsation in sdB stars based on our spectroscopic
measurements. On a more encouraging note, the results presented here will
provide a valuable resource for theoretical work into diffusion in sdB
stars. It will be important to study the diffusion of all iron group
and heavier elements individually in order to explain the trend we
have uncovered. We end then with a question and a challenge: how can we
find the reason some sdB pulsate and some don't if not by
spectroscopic means?

\begin{acknowledgements}
We thank the referee for constructive comments that have improved the
manuscript. We would like to thank  Michael Lemke for his support with the
LINFOR code and Norbert Przybilla for useful
discussions on the quality of atomic data. SJOT is supported by the
Deutsches Zentrum f\"ur Luft- und Raumfahrt (DLR) through grant
no.\ 50-OR-0202.
\end{acknowledgements}

\bibliographystyle{aa}

\end{document}